\documentstyle[11pt,epsfig]{article}
\newcounter{subequation}[equation]

\makeatletter 
\def\thesubequation{\theequation\@alph\c@subequation}
\def\@subeqnnum{{\rm (\thesubequation)}}
\def\slabel#1{\@bsphack\if@filesw {\let\thepage\relax
   \xdef\@gtempa{\write\@auxout{\string
      \newlabel{#1}{{\thesubequation}{\thepage}}}}}\@gtempa
   \if@nobreak \ifvmode\nobreak\fi\fi\fi\@esphack}
\def\subeqnarray{\stepcounter{equation}
\let\@currentlabel=\theequation\global\c@subequation\@ne
\global\@eqnswtrue
\global\@eqcnt\z@\tabskip\@centering\let\\=\@subeqncr
$$\halign to \displaywidth\bgroup\@eqnsel\hskip\@centering
  $\displaystyle\tabskip\z@{##}$&\global\@eqcnt\@ne
  \hskip 2\arraycolsep \hfil${##}$\hfil
  &\global\@eqcnt\tw@ \hskip 2\arraycolsep
  $\displaystyle\tabskip\z@{##}$\hfil
   \tabskip\@centering&\llap{##}\tabskip\z@\cr}
\def\endsubeqnarray{\@@subeqncr\egroup
                     $$\global\@ignoretrue}
\def\@subeqncr{{\ifnum0=`}\fi\@ifstar{\global\@eqpen\@M
    \@ysubeqncr}{\global\@eqpen\interdisplaylinepenalty \@ysubeqncr}}
\def\@ysubeqncr{\@ifnextchar [{\@xsubeqncr}{\@xsubeqncr[\z@]}}
\def\@xsubeqncr[#1]{\ifnum0=`{\fi}\@@subeqncr
   \noalign{\penalty\@eqpen\vskip\jot\vskip #1\relax}}
\def\@@subeqncr{\let\@tempa\relax
    \ifcase\@eqcnt \def\@tempa{& & &}\or \def\@tempa{& &}
      \else \def\@tempa{&}\fi
     \@tempa \if@eqnsw\@subeqnnum\refstepcounter{subequation}\fi
     \global\@eqnswtrue\global\@eqcnt\z@\cr}
\let\@ssubeqncr=\@subeqncr
\@namedef{subeqnarray*}{\def\@subeqncr{\nonumber\@ssubeqncr}\subeqnarray}
\@namedef{endsubeqnarray*}{\global\advance\c@equation\m@ne%
                           \nonumber\endsubeqnarray}

\makeatletter
\@addtoreset{equation}{section}
\makeatother
\renewcommand{\theequation}{\thesection.\arabic{equation}}
\def\dalemb#1#2{{\vbox{\hrule height .#2pt
        \hbox{\vrule width.#2pt height#1pt \kern#1pt
                \vrule width.#2pt}
        \hrule height.#2pt}}}

    \let\e=\epsilon
  \let\q=\theta  
  \let\n=\nu

\def\nn{\nonumber} \def\bd{\begin{document}} \def\ed{\end{document}}
\def\ds{\documentstyle} \let\fr=\frac \let\bl=\bigl \let\br=\bigr
\let\Br=\Bigr \let\Bl=\Bigl 
\let\bm=\bibitem
\let\na=\nabla
\let\pa=\partial \let\ov=\overline
\def\ie{{\it i.e.\ }} 
\newcommand{\be}{\begin{equation}} 
\newcommand{\ee}{\end{equation}} 
\def\ba{\begin{array}}
\def\ea{\end{array}}
\def\ft#1#2{{\textstyle{{\scriptstyle #1}\over {\scriptstyle #2}}}}
\def\fft#1#2{{#1 \over #2}}
\def\del{\partial}
\def\sst#1{{\scriptscriptstyle #1}}
\def\oneone{\rlap 1\mkern4mu{\rm l}}
\def\e7{E_{7(+7)}}
\def\td{\tilde}
\def\wtd{\widetilde}
\def\im{{\rm i}}
\def\bog{Bogomol'nyi\ }
\def\q{{\tilde q}}
\def\hast{{\hat\ast}}
\def\0{{\sst{(0)}}}
\def\1{{\sst{(1)}}}
\def\2{{\sst{(2)}}}
\def\3{{\sst{(3)}}}
\def\4{{\sst{(4)}}}
\def\5{{\sst{(5)}}}
\def\6{{\sst{(6)}}}
\def\7{{\sst{(7)}}}
\def\8{{\sst{(8)}}}
\def\n{{\sst{(n)}}}
\def\oo{{\"o}}
\def\hA{\hat{\cal A}}
\def\ns{{\sst {\rm NS}}}
\def\rr{{\sst {\rm RR}}}
\def\tH{{\widetilde H}}
\def\tB{{\widetilde B}}
\def\cA{{\cal A}}
\def\cF{{\cal F}}
\def\tF{{\wtd F}}
\def\Z{\rlap{\sf Z}\mkern3mu{\sf Z}}
\def\ep{{\epsilon}}
\def\IIA{{\rm IIA}}
\def\IIB{{\rm IIB}}
\def\ads{{\rm AdS}}
\def\R{\rlap{\rm I}\mkern3mu{\rm R}}
\def\mapright#1{\smash{\mathop{-\!\!\!-\!\!\!-\!\!\!-\!\!\!-\!\!\!
             \longrightarrow}\limits^{#1}}}

\def\Ei{{\hbox{Ei}}}
\def\Ci{{\hbox{Ci}}}
\def\Si{{\hbox{Si}}}

\newcommand{\ho}[1]{$\, ^{#1}$}
\newcommand{\hoch}[1]{$\, ^{#1}$}
\newcommand{\bea}{\begin{eqnarray}} 
\newcommand{\eea}{\end{eqnarray}} 
\newcommand{\ra}{\rightarrow}
\newcommand{\lra}{\longrightarrow}
\newcommand{\Lra}{\Leftrightarrow}
\newcommand{\aap}{\alpha^\prime}
\newcommand{\bp}{\tilde \beta^\prime}
\newcommand{\tr}{{\rm tr} }
\newcommand{\Tr}{{\rm Tr} } 
\newcommand{\NP}{Nucl. Phys. }
\newcommand{\tamphys}{\it Center for Theoretical Physics,
Texas A\&M University, College Station, TX 77843}
\newcommand{\upenn}{\it Dept. of Phys. and Astro., 
University of Pennsylvania,
Philadelphia, PA 19104}

\newcommand{\auth}{M. Cveti\v{c}, H. L\"{u} and J. F. V\'azquez-Poritz}

\begin{document}
\begin{flushright}
UPR/0878-T \\
February  2000\\
\hfill{\bf hep-th/0002128}\\
\end{flushright}

%\vspace{10pt}

\begin{center}

{\large {\bf  Absorption by Extremal D3-branes}
}

\vspace{20pt}

\auth

\vspace{10pt}
{\hoch{\dagger}\upenn}

\vspace{30pt}

\underline{ABSTRACT}
\end{center}

   The absorption in the extremal D3-brane background is studied for
a class of massless fields whose linear perturbations leave the
ten-dimensional background metric unperturbed, as well as the
minimally-coupled massive scalar.  We find that various fields have
the same absorption probability as that of the dilaton-axion system,
which is given exactly via the Mathieu equation.  We analyze the
features of the absorption cross-sections in terms of effective
Schr\"odinger potentials, conjecture a general form of the dual
effective potentials, and provide explicit numerical results for the
whole energy range.  As expected, all partial-wave absorption
probabilities tend to zero (one) at low (large) energies, and exhibit
an oscillatory pattern as a function of energy.  The equivalence of
absorption probabilities for various modes has implications for the
correlation functions on the field, including subleading contributions
on the field-theory side.  In particular, certain half-integer and
integer spin fields have identical absorption probabilities, thus
providing evidence that the corresponding operator pairs on the field
theory side belong to the same supermultiplets.

{\vfill\leftline{}\vfill
\vskip 10pt \footnoterule {\footnotesize \hoch{1} Research supported
in part by DOE grant DOE-FG02-95ER40893
\vskip  -12pt} \vskip   14pt
%{\footnotesize
%        \hoch{2}        Research supported in part by DOE 
%grant DOE-FG03-95ER40917 \vskip -12pt}  \vskip  14pt
}

\pagebreak
\setcounter{page}{1}

%\tableofcontents
%\addtocontents{toc}{\protect\setcounter{tocdepth}{2}}
%\newpage

\section{Introduction}

Scattering processes in the curved backgrounds of p-brane
configurations of M theory and string theory have been extensively
studied over the past few years. Motivation for these studies is the
fact that the low energy absorption cross-sections for different
fields yield information about the two-point correlation functions in
strongly coupled gauge theories via AdS/CFT correspondence
\cite{malda,gkp,wit}.

The extremal D3-brane background is of special interest, since the
correspondence there is to D=4 super Yang-Mills theory. By now, there
have been extensive studies of the scattering processes at
low-energies for the minimally-coupled scalar field (dilaton-axion)
from the gravity perspective, as well as the analyses of the
correlation functions on the field-theory side (see, for example,
\cite{kleb,GKT,ghkk,gubser,KRT,Friedman,Liu} and references therein),
thus leading to important insights into the AdS/CFT correspondence.
In particular, the supersymmetry constraints imply that the two-point
correlation functions do not get renormalized by higher-order
corrections on the field theory side, and a precise agreement between
low-energy absorption cross-sections for all the partial waves of the
dilaton field and the (weak coupling) field-theory calculations of the
corresponding correlation functions was obtained \cite{KRT}.

On the other hand, the scattering of other fields has been explored to
a lesser extent, and that only for low-energies (see, for example,
\cite{mathur,hoso} and references therein); nevertheless it is
expected that non-renormalization theorems on the field theory side
ensure a precise agreement between the low energy absorption
cross-sections and the corresponding weak coupling calculation of the
n-point correlation functions on the field-theory side (see, for
example, \cite{Friedman,Liu,dhoker} and references therein).
 
    This paper addresses several issues. We provide an analysis of the
absorption cross-section in the extremal D3-brane background for a
broad class of massless modes and for the whole energy range. In
particular, we uncover a pattern in the energy dependence of the
absorption cross-sections for both integer and half-integer spins;
certain half-integer and integer spin pairs have identical absorption
cross-sections, thus providing an evidence on the gravity side that
such pairs couple on the dual field theory side to the pairs of
operators forming supermultiplets of strongly coupled gauge theory.
 
      The paper is organized as follows.  In Section 2, we cast the
wave equations of various fields into Schr\"odinger form, and obtain
the effective Schr\"odinger potentials.  We show that effective
Schr\"odinger potentials for certain field are identical and hence the
absorption for these fields is the same. In other cases where
Schr\"odinger potentials are not the same, we argue that the different
potentials are dual and yield the same absorption probabilities, with
numerical results supporting these claims.  In Section 3, we obtain
numerical results for the absorption of a large class of fields for
the whole energy range in the extremal D3-brane background.  The
method for the numerical evaluation of the absorption probabilities is
given in Appendix A, while the calculation of the high energy
absorption cross-section in the geometrical optics limit is given in
Appendix B.

\section{Effective potentials}

The  D3-brane of the type IIB supergravity is given by
%%%%%%%
\bea
ds^2_{10}&=&H^{1/2} (-fdt^2 + dx_1^2 + dx_2^2 + dx_3^2) +
H^{1/2} (f^{-1} dr^2 + r^2\, d\Omega_5^2)\,,\nn\\
G_\5 &=& d^4x\wedge dH^{-1} + {*(d^4x\wedge dH^{1}})\,.\label{d3metric}
\eea
where 
\be
H=1 + {R^4\over r^4}, \ \ f=1- {{2m}\over r^4}.
\ee
%%%%%
Here $R$ specifies the D3-brane charge and $m$ is the non-extremality
parameter (defined for convenience as $m\equiv \mu
R^4/\sqrt{1+\mu}$). We shall primarily concentrate on the extreme
limit $m=\mu=0$. (See, however, Appendix B for the discussion of the
high energy limit of the absorption cross-section in the non-extreme
D3-background.)

The low energy absorption probabilities for various bosonic
linearly-excited massless fields under this background were obtained
in \cite{mathur}.  The low energy absorption probabilities for the
gravitino and the two-form field are given in \cite{hoso} and
\cite{raja}, respectively (and for massive minimally coupled modes in
\cite{mass}).  In the following subsections we study the absorption
probabilities for the whole energy range and uncover completely
parallel structures.  In particular, we shall cast the wave equation
for different modes into Schr\"odinger form and discuss the pattern of
the the Schr\"odinger potentials. We also provide a conjectured form
of the dual potentials which, in turn, yield the same absorption
probabilities. In the subsequent section we confirm the pattern with
numerical results.

\subsection{Dilaton-axion}

The axion and dilaton of the type IIB theory are decoupled from the
D3-brane. Thus, in  the D3-brane background, they satisfy the 
minimally-coupled scalar wave equation
%%%%
\be
\fft{1}{\sqrt{g}} \del_\mu \sqrt{g} g^{\mu\nu}\del_\nu\phi=0\,.
\ee
%%%%%
It follows from (\ref{d3metric}) that the radial wave equation of a
dilaton-axion in the spacetime of an extremal D3-brane is given by
%%%%%%%%
\be
\Big( \frac{1}{\rho^5} \frac{\partial}{\partial \rho} \rho^5 
\frac{\partial}{\partial \rho} + H - 
\frac{\ell (\ell + 4)}{\rho^2} \Big) \phi (\rho) = 0, \label{D3eqn}
\ee
%%%%%%%%%
where
%%%%%%%%
\be
H=1+\frac{e^4}{\rho^4}
\ee
%%%%%%%%
and $\ell=0,1,\ldots$ corresponds to the $\ell^{th}$ partial  wave.

The quantity $e$ and $\rho$ are dimensionless energy and radial
distance parameters: $e=\omega R$ and $\rho=\omega r$.   The leading
order and sub-leading order cross-sections of the minimally-coupled
scalar by the D3-brane background were obtained in \cite{kleb,ghkk}
by matching inner and outer solutions of the wave equations.  It was
observed in \cite{gubser} that if one performs the following change of
variables
%%%%%
\be
\rho=e\, Exp(-z)\,,\qquad \phi(r) = Exp(2z)\,\psi(r)\,,
\ee
%%%%%
the wave equation (\ref{D3eqn}) becomes
%%%%%
\be
\Big[\fft{\del^2}{\del z^2} + 2e\, \cosh(2z) - (\ell+2)^2\Big]\, 
\psi(z)=0\,,
\ee
%%%%%
which is precisely the modified Mathieu equation.  One can then
obtain analytically the absorption probability order by order in
terms of dimensionless energy $e$ \cite{gubser}.  

In this paper, we shall express the wave equation in Schr\"odinger form, and 
study the characteristics of the Schr\"odinger effective potential.  By
the substitution
%%%%%%%%
\be
\phi = \rho^{-5/2} \psi,
\ee
%%%%%%%%
we render (\ref{D3eqn}) into Schr\"odinger form
%%%%%%%%
\be
\big(\frac{\partial^2}{\partial \rho^2}-V_{\rm eff} \big)\psi = 0,
\label{eqdil}
\ee
%%%%%%%%
where 
%%%%%%%%
\be
V_{\rm eff}(\ell) = -H+\frac{(\ell+3/2)(\ell+5/2)}{\rho^2} \equiv
V_{\rm dilaton}(\ell).
\label{V}
\ee
%%%%%%%
Factors shared by the incident and outgoing parts of the wave
function cancel out when calculating the absorption probability. Thus,
the absorption probability of $\phi$ and $\psi$ are the same.

Technically, $V_{\rm eff}$ cannot be interpreted as an effective
potential, since it depends on the particle's incoming energy. 
It is straightforward to use a coordinate transformation to put
the equation in the standard Schr\"odinger form, where $V_{\rm eff}$
is independent on the energy.   However,
for our purposes of analyzing and comparing the form of the wave
equations for various fields, this is of no consequence.

Note that the first term in (\ref{V}) represents the spacetime geometry of
the extremal D3-brane, whereas the second term represents the angular
dynamics (partial modes) of the particles. We shall see presently that
some particles have effective potentials which contain terms mixed with
both ``geometrical'' and ``angular'' dynamical contributions.

\subsection{Antisymmetric tensor from 4-form}

For two free indices of the 4-form along $S^5$ and two free indices in
the remaining 5 directions, the radial wave equation for the
antisymmetric tensor derived from the 4-form is \cite{mathur}
%%%%%%
\be
\Big( \frac{1}{\rho} \frac{\partial}{\partial \rho} \rho
\frac{\partial}{\partial \rho} +H-\frac{(\ell+2)^2}{\rho^2} \Big) \phi
(\rho)=0, \label{4-form}
\ee
%%%%%%
where $\ell=1,2,\ldots$.  By the substitution
%%%%%%%
\be
\phi=\rho^{-1/2} \psi,
\ee
%%%%%%%
we render (\ref{4-form}) into Schr\"odinger form with
%%%%%%%
\be
V_{\rm 4-form}(\ell)=V_{\rm dilaton}(\ell).
\ee
%%%%%%%%

\subsection{Dilatino}

The radial wave equation for the dilatino on an extremal D3-brane was
found \cite{hoso} by inserting the following spherical wave
decomposition form for the dilatino field $\lambda$ into the covariant
Dirac equation:
%%%%%%%
\be
H^{\frac{1}{8}} \lambda=e^{-i \omega t} r^{-\frac{5}{2}} \Big( F(r)
\Psi_{-\ell}^{\pm} +i G(r) \big( \frac{\Gamma^{0} \Gamma^{i} x_{i}}{r}
\big) \Psi_{-\ell}^{\pm} \Big), 
\ee
%%%%%%%
where $\Gamma^i$ are field-independent gamma-matrices and
$i={4,\ldots,9}$ runs normal to the brane.  $\Psi_{-\ell}^{\pm}$ is the
eigenspinor of the total angular momentum with $\Sigma_{ij}
L_{ij}=-\ell$, $\Gamma^{0123}=\pm i$.

The spatial momenta tangential to the branes can be made to vanish via
the Lorentz transformations and the radial wave equations are obtained
\cite{hoso}:
%%%%%%%%
\bea
\omega H^{\frac{1}{2}} F+\Big(
\frac{d}{dr}+\frac{\ell+5/2}{r} \pm \frac{1}{4}(\ln H)' \Big) G&=&0
\label{f}\\
-\omega H^{\frac{1}{2}} G+\Big(
\frac{d}{dr}-\frac{\ell+5/2}{r} \mp \frac{1}{4}(\ln H)' \Big) F&=&0
\label{ff}
\eea
%%%%%%%%
Decoupling (\ref{f}) and (\ref{ff}) yields second-order differential
equations for $F$ and $G$. 

Let us first consider the case of positive eigenvalue, {\it
i.e.}, $\Gamma^{0123}=+i$.  In this case, the second-order wave
equation for $F$ can be cast into Schr\"odinger form by the
substitution $F=H^{1/4}\, \psi$, giving rise to the effective
Schr\"odinger potential
%%%%%%%%
\be
V^F_{\rm +dilatino}(\ell)= V_{\rm dilaton}(\ell)\,.\label{dilatino1}
\ee
%%%%%%%%
Thus, the absorption probability for $F$ is identically the same
as that for the dilaton-axion.  The wave equation for $G$ can also 
be cast into Schr\"odinger form by the substitution
$G=H^{1/4}\, \psi$.
The corresponding Schr\"odinger potential, on the other hand, takes a 
different form, given by
%%%%%%%%
\be
V^G_{\rm +dilatino}(\ell)= V_{\rm dilaton}(\ell) + \frac{-3e^8 -2e^8
\ell-10e^4
\rho^4 +5\rho^8 +2\rho^8 \ell}{H^2 \rho^{10}}\,.\label{dilatino2}
\ee
%%%%%%%%
Thus, we see that the effective potentials for the two components $F$
and $G$ are quite different.  However, we expect that these two
potentials, although different, yield the same absorption probability:
they form a dual pair of potentials.  While it is not clear to us how to
present a rigorous analytical proof, our numerical calculation (in section
4) confirm that, indeed, they yield the same absorption probability.
 
Similar results are obtained for the the negative eigenvalue
solutions, {\it i.e.}, $\Gamma^{0123}=-i$:
%%%%%%%%
\bea
V^G_{\rm -dilatino}(\ell)&=& V_{\rm dilaton}(\ell +1)\,,\nn\\
V^F_{\rm -dilatino}(\ell)&=& V_{\rm dilaton}(\ell +1) + 
\frac{-5\rho^8 -2\rho^8
\ell-10e^4 \rho^4 +7e^8 +2e^8 \ell}{H^2 \rho^{10}}.
\eea
%%%%%%%%
Again, the numerical results in section 4 indicate that the
above two potentials yield the same absorption probabilities and hence
form a dual pair.

        In the  above discussion of the dilatino scattering equation, 
we  encountered three different potentials, namely 
%%%%%%
\be
V^G_{\rm +dilatino}(\ell)\mapright{\rm dual}
V^F_{\rm +dilatino}(\ell)=V^G_{\rm -dilatino}(\ell-1)\mapright{\rm dual}
V^F_{\rm -dilatino}(\ell-1)
\ee
which all yield the same absorption probability.

   The above structure of the dual potentials can be cast in a more
general form.  Namely, the above dual potential pairs for the
dilatino can be cast into the following form:
%%%%%%%
\be
V(\ell)=-H+\frac{(2\ell+\alpha)(2\ell+\alpha \pm 2)}{4\rho^2}, \label{form}
\ee
%%%%%%%
and
%%%%%%%
\be
V^{\rm dual}(\ell)=V(\ell)+ \frac{\pm[(2\ell+\alpha \pm 2)e^8-
(2\ell +\alpha)\rho^8]-10e^4 \rho^4}{\rho^{10} H^2},
\label{dual}
\ee
%%%%%%%
where $\alpha=5$ and we use in the $\pm$ sign in (\ref{dual}) for the
$\mp$ eigenvalue, respectively. We have found numerically that
(\ref{form}) and (\ref{dual}) are dual potentials for integer values
of $\alpha$. In particular, for odd values of $\alpha$, (\ref{form})
can be identified with $V_{\rm dilaton}(\ell+(\alpha-4 \pm 1)/2)$, in
which case the absorption can be found analytically, since the wave
equation is that of Mathieu equation.

Thus, in all the subsequent examples when the potential is of the
form (\ref{dual}), we are now able to identify a dual potential
(\ref{form}) of a simpler form with a corresponding wave equation
which can be solved analytically.

\subsection{Scalar from the two-form}

For the free indices of the two-form taken to lie along the $S^5$, the
radial wave equation is \cite{mathur}
%%%%%%%%
\be
\Big( \frac{H}{\rho} \frac{\partial}{\partial \rho} \frac{\rho}{H} 
\frac{\partial}{\partial \rho} + H - \frac{(\ell+2)^2}{\rho^2} 
\mp \frac{4e^4}{H \rho^6} (\ell+2) \Big) \phi
(\rho) = 0, \label{scalar}
\ee
%%%%%%%%%
where again $\ell=1,2,\ldots$ correspond to the $\ell^{th}$ partial
wave.  The sign $\pm$ corresponds to the sign in the spherical
harmonics involved in the partial wave expansion.

By the substitution
%%%%%%%%
\be
\phi = \rho^{-1/2} H^{1/2} \psi,
\ee
%%%%%%%%
we render (\ref{scalar}) into Schr\"odinger form.

      For the positive eigenvalue, the effective potential is of the
form given by (\ref{dual}), with $\alpha=5$ and a positive sign. This
is, in fact, the same as that for the dilatino with negative eigenvalue.
Thus, the dual potential is
%%%%%%%%
\be
V_{\rm +scalar}(\ell)\mapright{\rm dual} V_{\rm dilaton}(\ell +1).
\ee
%%%%%%%%
For the negative eigenvalue, the effective potential is of the form
given by (\ref{dual}), with $\alpha=3$ and a negative sign.
Thus, the dual potential is
%%%%%%%%
\be
V_{\rm -scalar}(\ell)\mapright{\rm dual} V_{\rm dilaton}(\ell -1).
\ee
%%%%%%%%

\subsection{Two-form from the antisymmetric tensor}

The equations for two-form perturbations polarized along the
D3-brane are coupled. For s-wave perturbations, they can be
decoupled \cite{mathur,raja}, and the radial wave equation is
%%%%%%%%
\be
\Big( \frac{1}{\rho^5 H} \frac{\partial}{\partial \rho} \rho^5 H 
\frac{\partial}{\partial \rho} + H - 
\frac{16 e^8}{\rho^{10} H^2} \Big) \phi (\rho) = 0, \label{2-form}
\ee
%%%%%%%%%
where $\ell=0,1,\ldots.$   By the substitution
%%%%%%%%
\be
\phi = \rho^{-5/2} H^{-1/2} \psi,
\ee
%%%%%%%%
we render (\ref{2-form}) into Schr\"odinger form given by 
(\ref{dual}), with $\alpha=5$, $\ell=0$ and the positive sign.
Thus, the dual potential is

%%%%%%%%
\be
V_{\rm 2-form}(\ell=0)\mapright{\rm dual} V_{\rm dilaton}(\ell=1).
\ee
%%%%%%%%

\subsection{Vector from the two-form}

We now consider one free index of the two-form along $S^5$ and one
free index in the remaining 5 directions. For the tangential components of
this vector field, the radial wave equation is
\cite{mathur}
%%%%%%%%
\be
\Big( \frac{1}{\rho^3} \frac{\partial}{\partial \rho} \rho^3 
\frac{\partial}{\partial \rho} + H - 
\frac{(\ell+1)(\ell+3)}{\rho^2} \Big) \phi (\rho) = 0, \label{tang}
\ee
%%%%%%%%%
where $\ell=1,2,\ldots.$   By the substitution
%%%%%%%%
\be
\phi = \rho^{-3/2} \psi,
\ee
%%%%%%%%
we render (\ref{tang}) into Schr\"odinger form with
%%%%%%%
\be
V_{\rm tangential-vector}(\ell)=V_{\rm dilaton}(\ell).
\ee
%%%%%%%%%
The radial $a_r$ and time-like  $a_0$ components of the vector
field are determined by the following coupled first order differential
equations \cite{mathur}:
%%%%%%%
\be
i \frac{\partial}{\partial \rho} a_o = 
\big[ 1-\frac{(\ell+1)(\ell+3)}{\rho^2 H}
\big] a_r \label{a}
\ee
%%%%%%%
and
%%%%%%%
\be
i \frac{1}{\rho} \frac{\partial}{\partial \rho} 
\big( \frac{\rho}{H} a_r \big) =
a_o, \label{aa}
\ee
%%%%%%
where $\ell=1,2,\ldots.$ Eqs. (\ref{a}) and (\ref{aa}) can be
decoupled and the wave equation for $a_r$ is
%%%%%%%%
\be
\Big( H \frac{\partial}{\partial \rho} \frac{1}{\rho} 
\frac{\partial}{\partial \rho} \frac{\rho}{H} + H - 
\frac{(\ell+1)(\ell+3)}{\rho^2} \Big) a_r = 0. \label{ar}
\ee
%%%%%%%%%
By the substitution
%%%%%%%%
\be
a_r = \rho^{-1/2} H \psi,
\ee
%%%%%%%%
we render (\ref{ar}) into Schr\"odinger form with
%%%%%%%
\be
V_{\rm radial-vector}(\ell)=V_{\rm dilaton}(\ell).
\ee
%%%%%%%
The wave equation for $a_0$ is
%%%%%%
\be
\Big( \frac{1}{\rho} \frac{\partial}{\partial \rho}
\frac{\rho^3}{\rho^2 H-(\ell+1)(\ell+3)} \frac{\partial}{\partial
\rho} +1 \Big) a_0 =0. \label{a0}
\ee
%%%%%%%
By the substitution
%%%%%%%
\be
a_0=\rho^{-3/2} \sqrt{\rho^2 H-(\ell+1)(\ell+3)} \psi,
\ee
%%%%%%%%
we render (\ref{a0}) into Schr\"odinger form with 
%%%%%%%
\be
V_{\rm 0-vector}(\ell)=V_{\rm dilaton}(\ell)+\frac{3e^8 -10e^4 \rho^4
+4\rho^6 (\ell+1) (\ell+3)-\rho^8}{\rho^{10} \big(
H-\frac{(\ell+1)(\ell+3)}{\rho^2} \big)^2}.
\ee
%%%%%%%
Note that this effective potential is nonsingular only for $e^2 >
(\ell+1)(\ell+3)/2$. We have numerically confirmed that $V_{\rm
0-vector}(\ell)$ is dual to $V_{\rm dilaton}(\ell)$. Working directly
from Eqs. (\ref{a})-(\ref{aa}), we have numerically confirmed that
$a_0$ shares the same absorption probability with the dilaton-axion,
for all values of $e$.

\section{Qualitative features of absorption by the extremal D3-brane}

We have obtained numerical absorption probabilities by a method
described in Appendix A. Fig. 1 shows the s-wave absorption
probabilities for all the particles that we have studied on an
extremal D3-brane vs. the energy of incoming particles measured in
dimensionless units of $e \equiv \omega R$. %%%%%%%%
\begin{figure}
   \epsfxsize=4.0in
   \centerline{\epsffile{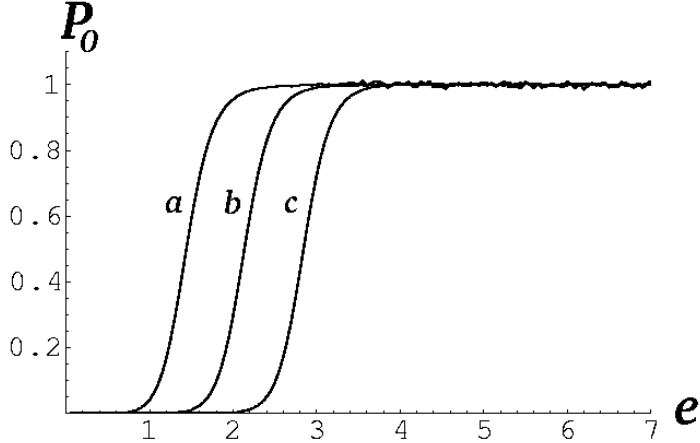}}
   \caption[FIG. \arabic{figure}.]{s-wave absorption
probabilities of various particles on an extremal D3-brane}
\end{figure}
%%%%%%%%
(a) is the s-wave absorption probability of the dilaton-axion, dilatino
with positive total angular momentum eigenvalue and scalar from the two-form 
with a negative sign in the spherical harmonic. (b) is the s-wave
absorption probability of the dilatino with negative total angular
momentum eigenvalue, two-form from the antisymmetric tensor,
antisymmetric tensor from the four-form and the longitudinal and
tangential components of the vector from the two-form. (c) is the s-wave
absorption probability of the scalar from the two-form with a positive
sign in the spherical harmonic. These absorption probabilities are simply
related by $\ell \rightarrow \ell \pm 1$. This structure is suggestive of
a particle supermultiplet structure-- namely, different multiplets.

This demonstrates the similar and surprisingly simple
structure of absorption probabilities between the various particles,
which is not apparent from previous analytical low-energy absorption
probabilities. These numerical results also support the idea of dual
potentials. 

Fig. 2 shows the partial absorption probabilities for a dilaton-axion on
an extremal D3-brane vs. the energy of incoming particles.
%%%%%%%%
\begin{figure}
   \epsfxsize=4.0in
   \centerline{\epsffile{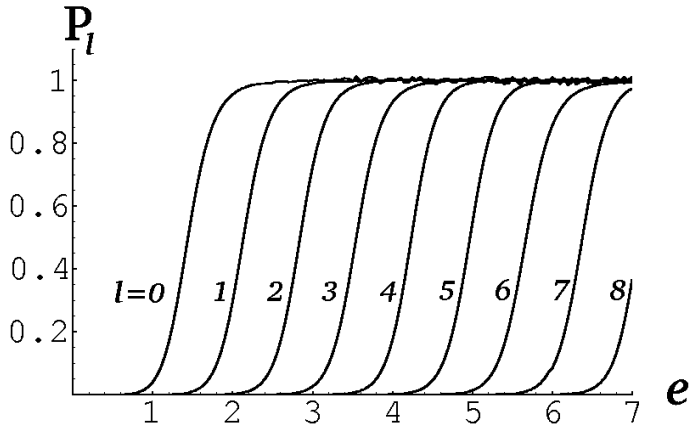}}
   \caption[FIG. \arabic{figure}.]{Partial absorption probabilities for a
dilaton-axion on an extremal D3-brane}
\end{figure}
%%%%%%%%
There is no absorption at zero energy and total absorption is approached
at high energy. Thus, the high energy absorption cross-section is found by
setting $P=1$ in (\ref{prob}):
%%%%%%
\be
\sigma^{(\ell)}=\frac{8 \pi^2
(\ell+3)(\ell+2)^2(\ell+1)}{3 (\omega R)^2}
\ee
%%%%%%%%
In fact, for all types of waves absorbed by all branes and black holes
studied thus far \cite{mynotes}, total partial wave absorption occurs at
high energy. It is conjectured that this is a general property of
absorption for all objects that have an event horizon. Results for objects
other than D3-branes will be published shortly by the present authors. 

The partial absorption probabilities of a massive minimally-coupled scalar
on an extremal D3-brane vs. the energy of the incoming particles is shown
in Fig. 3. 
%%%%%%%%
\begin{figure}
   \epsfxsize=4.0in
   \centerline{\epsffile{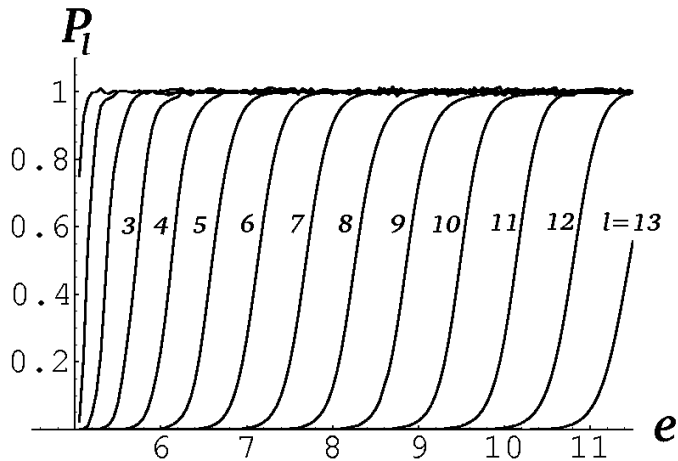}}
   \caption[FIG. \arabic{figure}.]{Partial absorption probabilities for
a massive minimally-coupled scalar on an extremal D3-brane}
\end{figure}
%%%%%%%%
As the mass of the scalar is increased, the partial absorption
probabilities occur at lower energies. Physically, there is greater
absorption at low energy due to the additional gravitational
attraction that is present from the nonzero mass.  The low energy
absorption probability for this case were obtained in \cite{mass}.

Fig. 4 shows the numerical s-wave absorption cross-section for a
dilaton-axion on an extremal D3-brane (continuous line), super-imposed
with previously obtained low energy semi-analytical results (short
dashes) \cite{gubser}, as well as high energy total absorption (long
dashes) vs. the energy of incoming particles. Throughout the rest
of this paper, absorption cross-sections are plotted in units of $R^5$.
%%%%%%%%
\begin{figure}
   \epsfxsize=4.0in
   \centerline{\epsffile{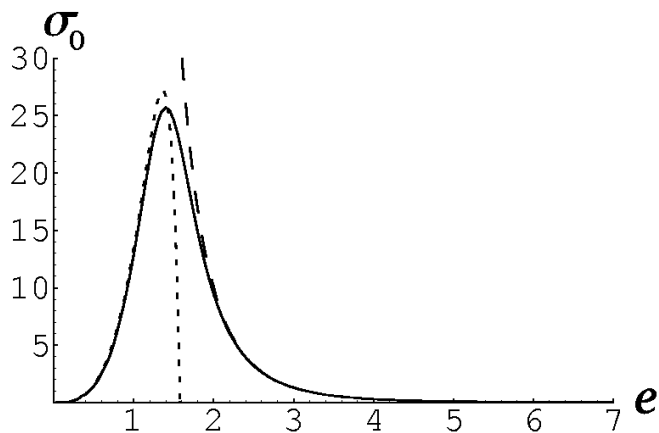}}
   \caption[FIG. \arabic{figure}.]{s-wave absorption
cross-section for a dilaton-axion on an extremal D3-brane}
\end{figure}
%%%%%%%%
The resonance roughly corresponds to the region in which there is a
transition from zero absorption to total absorption. 

As evident from Fig. 5, the energies at which there is a resonance in the  
partial absorption cross-section are proportional to the partial-wave
number. 
%%%%%%%%
\begin{figure}
   \epsfxsize=4.0in
   \centerline{\epsffile{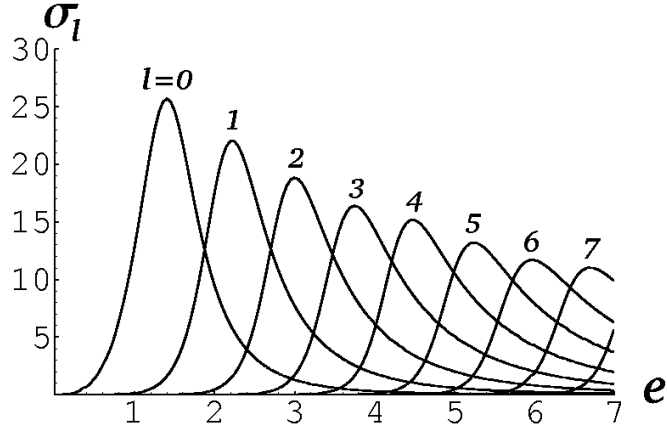}}
   \caption[FIG. \arabic{figure}.]{Partial absorption cross-sections for a
dilaton-axion on an extremal D3-brane}
\end{figure}
%%%%%%%%
Also, the magnitude of the peak of each partial absorption cross-section
decreases with the partial-wave number. These characteristics seem
reasonable if one considers particle dynamics; in terms of radial motion,
rotational kinetic energy counteracts gravitational attraction. 

In Fig. 6, we plot the partial-wave effective potentials vs. radial
distance (in dimensionless units) for a dilaton-axion on an extremal
D3-brane. These are plotted at the energies of the maxima of the 
corresponding partial-wave absorption cross-sections. For the s-wave 
absorption cross-section, the maximum is at $e=1.4$. The maxima of higher 
partial-wave absorption cross-sections are separated by energy gaps of
approximately $\Delta e=.75$, with increasing $\ell$.

%%%%%%%%
\begin{figure}
   \epsfxsize=4.0in
   \centerline{\epsffile{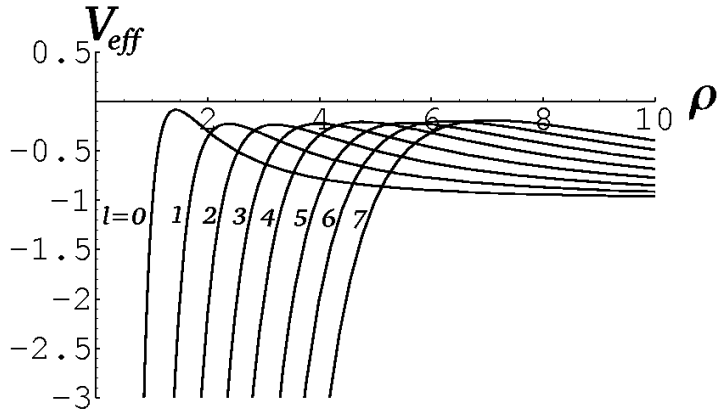}}
   \caption[FIG. \arabic{figure}.]{Partial effective potentials absorption
for a dilaton-axion on an extremal D3-brane}
\end{figure}
%%%%%%%%
As can be seen, the incoming particle must penetrate an effective
barrier in order to be absorbed by the D3-brane. Once absorbed, the
waves inhabit quasi-bound states until they quantum tunnel to
asymptotically flat spacetime. As is shown, the heights of the
partial-wave effective potential barriers at the energies of the
maxima of the partial-wave absorption cross-sections are roughly equal,
which partly explains the similar structure of the partial absorption
probabilities of different partial-wave numbers. Thus, the decreasing
maximum values of the partial-wave absorption cross-sections with
increasing $\ell$ arises purely as a result of the phase-factors.

The superposition of maxima of partial-wave absorption cross-section
leads to the oscillatory character of the total absorption
cross-section with respect to the energy of the incoming
particles. This is shown in Fig. 7 for the case of the dilaton-axion,
dilatino and scalar from the two-form for comparison.
%%%%%%%%
\begin{figure}
   \epsfxsize=4.0in
   \centerline{\epsffile{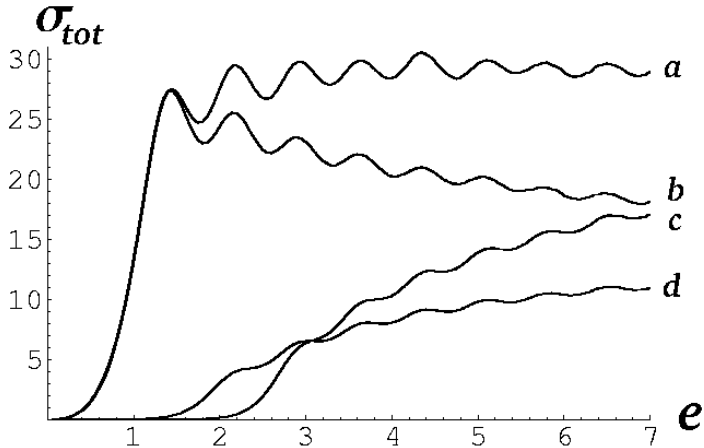}}
   \caption[FIG. \arabic{figure}.]{Total absorption cross-sections for
the dilaton-axion, dilatino and scalar from two-form on an extremal
D3-brane}
\end{figure}
%%%%%%%%
(a) is the total absorption cross-section of the dilaton-axion on an
extremal D3-brane, (b) is that of the dilatino with positive total
angular momentum eigenvalue, (c) is that of the scalar from the
two-form with positive sign in the spherical harmonic, and (d) is that
of the dilatino with negative total angular momentum eigenvalue.

The amplitude of oscillation decreases exponentially with energy. For
the dilaton-axion, the total absorption cross-section converges to the
geometrical optics limit at high energy, which we have calculated in a
previous section.  The oscillatory behavior is shared by all total
absorption cross-sections that have been studied as of this time
\cite{mynotes}, for various particles in various spacetime
backgrounds. The oscillatory structure of the absorption
cross-section of a scalar on a Schwarzschild black hole has been noted
by Sanchez \cite{sanchez}.

It is interesting to note that extinction cross-sections which arise
in the field of optics have similar oscillatory properties \cite{hulst}.

Fig. 8 shows the total absorption cross-section of the scalar from the
two-form with negative total angular momentum eigenvalue.
%%%%%%%%
\begin{figure}
   \epsfxsize=4.0in
   \centerline{\epsffile{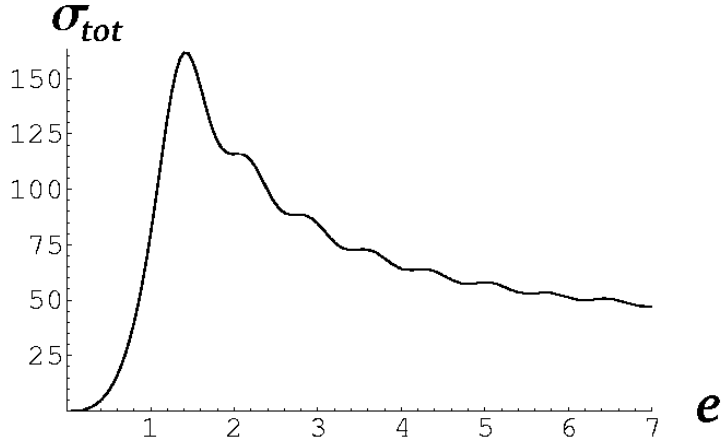}}
   \caption[FIG. \arabic{figure}.]{Total absorption cross-section for
the scalar from two-form with negative total angular momentum eigenvalue}
\end{figure}
%%%%%%%%

As already noted and shown in Fig. 3, as the mass of the
minimally-coupled scalar is increased, the partial-wave absorption
probabilities occur at lower energies. This causes the drastic
qualitative difference between the massless and massive scalar cases
at low energy, i.e., the divergent total absorption cross-section at
zero energy for the massive scalar, as is shown in Fig. 9.
%%%%%%%%
\begin{figure}
   \epsfxsize=4.0in
   \centerline{\epsffile{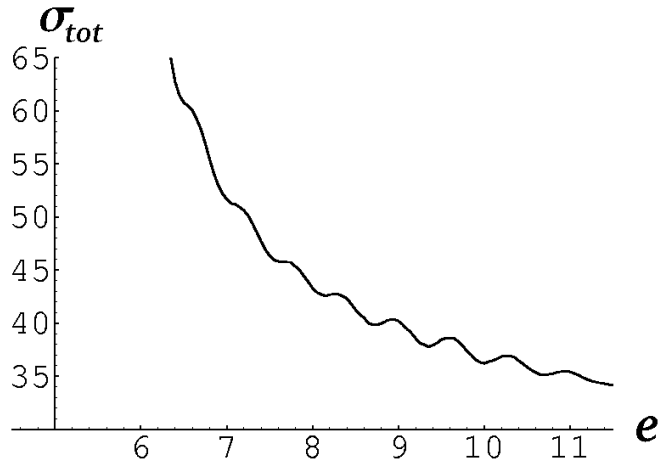}}
   \caption[FIG. \arabic{figure}.]{Total absorption cross-section for
a massive minimally-coupled scalar on an extremal D3-brane}
\end{figure}
%%%%%%%%

\section*{Acknowledgments}

We would like to thank C. Pope for useful discussions and J.V.P. would
like to thank A. Maarouf for his assistance in formating the plots.

\appendix

\section{Outline of Numerical Method}

We will now outline the numerical method for finding the absorption
cross-section. We will use the well-studied case of the dilaton-axion on
an extremal D3-brane. In this case, the radial wave equation is given by 
(\ref{eqdil}) and (\ref{V}).
We take the wave close to the horizon, at $\rho = .01$, to be purely
incoming:
%%%%%%%%%
\be
\psi (\rho) = \rho \exp{\big(\frac{i (\omega
R)^2}{\rho}\big)}
\label{bndry}
\ee
%%%%%%%%
The solution in the far region is:
\be
\psi (\rho) = A_{in} \exp{(i \rho)} + A_{out} \exp{(- i \rho)}.
\label{far}
\ee
%%%%%%% 

We use Mathematica to numerically integrate (\ref{eqdil}) with the
boundary condition given by (\ref{bndry}). At $\rho=45$, we match the
result with (\ref{far}) to find $A_{in}$ and $A_{out}$.
The absorption probability is given by
%%%%%%%
\be
P = 1 - |\frac{A_{out}}{A_{in}}|^2.
\ee
The absorption cross-section for a scalar is
%%%%%%%
\be
\sigma^{(\ell)}=\frac{8 \pi^2
(\ell+3)(\ell+2)^2(\ell+1)}{3 (\omega R)^2} P^{(\ell)} \label{prob}
\ee
%%%%%%%%
Numerical integration for other types of particles incident on other
objects with event-horizons is straight-forward. 

\section{High-energy absorption cross-section for a
dilaton-axion on a nonextremal D3-brane}

In this appendix, we obtain the analytical high energy absorption
cross-section for a dilaton-axion in a {\it nonextremal}
D3-brane background, by employing the geometrical optics limit
for the classical motion of a particle \cite{teuk}.

For a nonextremal D3-brane, the metric takes the form (\ref{d3metric}).
The classical Lagrangian for a particle  is of the form:
%%%%%%%%%
\be
L=\frac{1}{2} g_{\alpha \beta} \dot{x}^{\alpha} \dot{x}^{\beta}.
\ee
%%%%%%%%%
$\dot{x}^{\alpha} \equiv dx^{\alpha}/d\lambda$, where $\lambda$ is an
affine parameter.  The Euler-Lagrange equations are
%%%%%%%%%%
\be
\frac{d}{d\lambda} \big( \frac{\partial L}{\partial \dot{x}^{\alpha}}
\big)
- \frac{\partial L}{\partial x^{\alpha}} = 0.
\ee
%%%%%%%%%
The equation of motion for $\theta$ is
%%%%%%%%%
\be
\frac{d}{d\lambda} (H^{1/2} r^2 \dot{\theta}) = H^{1/2} r^2 \sin \theta
\cos \theta (\dot{\phi_3}^2 - \dot{\phi}^2 - \cos^2 \phi \dot{\phi_1}^2 -
\sin^2 \phi \dot{\phi_2}^2). \label{theta}
\ee
%%%%%%%%%
The solution of (\ref{theta}) is $\theta = \pi /2$ and $\dot{\theta} = 0$.
The equations of motion for $\phi_3$ and $t$ are of the form
%%%%%%%%
\be
\frac{d}{d\lambda} (H^{1/2} r^2 \dot{\phi_3} )=0 \label{phi}
\ee
%%%%%%%%
and
%%%%%%%%
\be
\frac{d}{d\lambda} (H^{-1/2} f \dot{t} )=0. \label{t}
\ee
%%%%%%%%
(\ref{phi}) and (\ref{t}) each imply a constant of motion:
%%%%%%%
\be
H^{1/2} r^2 \dot{\phi_3} = {\rm constant} \equiv \ell
\ee
%%%%%%
and
%%%%%%%
\be
H^{-1/2} f \dot{t} = {\rm constant} \equiv E,
\ee
%%%%%%%
where $\ell$ and $E$ are interpreted as the angular momentum and energy of
the particle, respectively.

Also, since the particle  scatters only in a direction transverse to the
D3-brane ({\it i.e.}, it does not  travel along the D3-brane), $\dot{x_i} =
0$, for $i = 1,2,3$.  Substituting our solutions for $\theta$ and 
$\dot{x_i}$ into the Lagrangian yields
%%%%%%%%%
\be
2L = - H^{-1/2} f \dot{t}^2 + H^{1/2} f^{-1} \dot{r}^2 + H^{1/2} r^2
\dot{\phi_3}^2.
\ee
%%%%%%%%
Instead of finding a rather complicated equation of motion for $r$, we use
the fact that, for a massless particle, 
%%%%%%%%
\be
2L = g_{\alpha \beta} p^{\alpha} p^{\beta} = 0,
\ee
%%%%%%%
where $p^{\alpha} = \dot{x}^{\alpha}$.
Substituting our results for $\dot{\phi_3}$ and $\dot{t}$ into the
previous equation yields
%%%%%%%%
\be
\dot{r}^2 = E^2 - \frac{\ell^2}{r^2} \frac{f}{H}.
\ee
%%%%%%%%
We introduce a new parameter, $\lambda^{\prime} \equiv \ell \lambda$, so
that
%%%%%%%
\be
\big( \frac{dr}{d\lambda^{\prime}} \big) ^2 = \frac{1}{b^2} -
V_{\rm effective},
\ee
%%%%%%%
where $b \equiv \ell /E$ is the impact parameter and
%%%%%%%
\be
V_{\rm effective} = \frac{1}{r^2} \frac{f}{H}.
\ee
%%%%%%%
The absorption cross-section for particles  at high energy  can 
be obtained by determining  the classical trajectory  of the scattered 
 particle and  using the optical limit result:
%%%%%%%%%
\be
\sigma_{\rm abs}=\frac{8}{15} \pi^2 b_{\rm crit}^5 ,\nn\\
\ee
%%%%%%%%%
where the  critical impact parameter separating absorption from scattering
orbits is given by $1/b_{\rm crit}^2 = V_{\rm maximum}$. Thus,
%%%%%%%%%
\bea
\sigma_{\rm abs} &=&\frac{8}{15} \pi^2 \Big(\frac{1}{2}R^4+3m+\frac{1}{2}
\sqrt{(R^4+6m)^2+8mR^4}
\Big)^{5/4} \times \nn\\
&=&\Big( \frac{\frac{3}{2}R^4+3m+\frac{1}{2} \sqrt{(R^4+6m)^2
+8mR^4}}{\frac{1}{2}R^4+m+\frac{1}{2} \sqrt{(R^4+6m)^2+8mR^4}} \Big)^{5/2}.
\eea
%%%%%%%%%
In the extremal limit, i.e. $m=0$, this result reduces to:
%%%%%%%%%
\be
\sigma_{\rm abs} = \frac{32 \sqrt{2}}{15} \pi^2 R^5.
\ee
%%%%%%%%%%

\end{document}